\documentclass[12pt]{article}

\usepackage{graphics}
\usepackage{graphicx}
\usepackage{amssymb}
\usepackage{amsmath}
\usepackage{amsfonts}

\newif\ifFIG
\FIGtrue

\begin{document}

\begin{titlepage}

\begin{center}

{\Large \bf  
{A Theoretical Study on a Reaction of \\ 
Iron(III) Hydroxide with Boron Trichloride \\
 by Ab Initio Calculation}
}

\vskip .45in

{\large
Kazuhide Ichikawa$^1$, Toshiyuki Myoraku$^1$, Akinori Fukushima$^1$, Yoshio~Ishihara$^2$, Ryuichiro Isaki$^2$, Toshio Takeguchi$^2$\\
 and Akitomo Tachibana$^{*,1}$}

\vskip .45in

{\em
$^1$Department of Micro Engineering, Kyoto University, Kyoto 606-8501, Japan \\
$^2$Taiyo Nippon Sanso Corporation, Tokyo 142-8558, Japan
}

\vskip .45in
{\tt E-mail: akitomo@scl.kyoto-u.ac.jp}

\end{center}

\vskip .4in

\begin{abstract}
We investigate a reaction of boron trichloride (BCl$_3$) with iron(III) hydroxide (Fe(OH)$_3$) by ab initio quantum chemical calculation as a simple model for a reaction of iron impurities in BCl$_3$ gas. We also examine a reaction with water. 
We find that compounds such as ${\rm Fe(Cl)(OBCl_2)_2(OHBCl_2)}$ and ${\rm Fe(Cl)_2(OBCl_2)(OHBCl_2)}$ are formed while producing HCl and reaction paths to them are revealed. We also analyze the stabilization mechanism of these paths using newly-developed interaction energy density derived from electronic stress tensor in the framework of the Regional DFT (Density Functional Theory) and Rigged QED (Quantum ElectroDynamics).
\end{abstract}

\end{titlepage}

\setcounter{page}{1}

\section{Introduction} \label{sec:intro}

In this paper, we investigate a reaction of boron trichloride (BCl$_3$) with iron(III) hydroxide (Fe(OH)$_3$) by ab initio quantum chemical calculation. The purpose of the paper is two fold. One is to present how Fe(OH)$_3$ reacts in BCl$_3$ gas which could be relevant to an industrial process. Secondly, by applying newly-developed interaction energy density concept to the reaction, we would like to test its validity in particular regarding its ability to describe stabilization through chemical reaction. 

Let us start from describing some industrial background. Boron trichloride 
 is one of the semiconductor gases, which is used, for example, as a preferential plasma etching gas for aluminum and a source of boron for p-type doping 
 in the process of chemical vapor deposition \cite{Herner2004,Kunii2004,Domaracka2005}.
 High purity is required for semiconductor gases to be used in the production process of recent high integrated device and thin insulating film. Required impurity level in recent years has been lower and lower. In the future, it is expected to be parts-per-trillion (ppt) level \cite{Briesacher1991}. To achieve lower impurity level, it is necessary to remove all causes which might contaminate the gases in the whole process including gas transportation. As examples of concerned impurities, there are water and metal. Once water or metallic impurities are mixed in BCl$_3$, they can deteriorate the product performance and the process yield. Generally, to remove impurities from ultra high purity gases, ceramic or synthetic resin filters are used. However, metallic impurities in boron containing gases, including BCl$_3$ gas, are not removed well through these types of filters \cite{JPPA2008} and reason for this is not known. It may mean that metallic impurities in boron containing gases have structures which easily pass through these filters as the result of the interaction between gases and metal. So far, methods for removing impurities in high purity gases have been discussed in the literatures, but structures or states of impurities have not been studied. In particular, it is difficult to directly observe the structure of impurities in gas phase. Therefore, we consider that it is worthwhile to investigate the molecular states of impurities by a computational method as a precursory study toward establishing a more effective method for removing impurities.

 Among several metallic impurities, since most abundant one is iron, we focus on iron-including impurities. We can think of several possible sources for iron impurities. The prime suspect is rust from the welded spots in the ductwork. Although the ductwork for semiconductor gases has corrosion resistance, the welded spots are relatively weak against halogen-contained gases, especially, under the presence of water \cite{Ishihara1994}. Note that at most a few parts-per-million (ppm) of water is mixed in the ductwork.

As a first step to guess how iron impurities react with BCl$_3$ gas, we consider a reaction of Fe(OH)$_3$ with BCl$_3$. The first reason why we pick up this iron compound is that we would like to see a reaction with a hydroxyl function since hydrochloric (HCl) gas is known to exist in BCl$_3$ gas. Such HCl gas may come from the reaction with the iron impurities. The second reason is that rust mentioned above is likely to be in the form of goethite ($\alpha$-FeO(OH)) \cite{Kim2005,Suzuki2005,Boily2008}, and its monohydrated form can be described as Fe(OH)$_3$.
 
This paper is organized as follows. In the next section, we briefly explain our quantum chemical computation method. We also describe our analysis method based on the Regional DFT (Density Functional Theory) and the Rigged QED (Quantum ElectroDynamics), and in particular we define the interaction energy density. In Sec.~\ref{sec:results}, we show our results on the reaction of BCl$_3$ with water and that with Fe(OH)$_3$. The final section is devoted to our conclusion. 
 
\section{Calculation Methods} \label{sec:calc}
\subsection{Ab initio electronic structure calculation}

We perform ab initio quantum chemical calculation for several chemical reactions using density functional theory. In this calculation, we adopt the Lee-Yang-Parr (LYP) \cite{Lee1988} gradient-corrected functionals for the correlation interaction, and it is employed with Becke's hybrid three parameters \cite{Becke1993} for generalized-gradient-approximation (GGA) exchange-correlation functions (B3LYP). We employ 6-311G* basis set for Fe, B, O and Cl, which consists of all electron basis set by Wachters-Hay \cite{Wachters1970,Hay1977}, and 6-311G** basis set for H. This calculation shows reasonable results for high-spin states of iron hydroxide clusters. Analytical vibrational frequencies are obtained to calculate the zero-point energy (ZPE) correction.
In this work, the geometric optimized structures and the electronic structures of each cluster model are calculated by {\sc{Gaussian03}} program package \cite{Gaussian03}. The part of visualization in this paper is done using MOLDEN\cite{MOLDEN} and VMD\cite{VMD} softwares. 

\subsection{Interaction energy density analysis}
In the following section, we use newly-developed interaction energy density in our laboratory to analyze how and in which part of molecules are (de-)stabilized during the chemical process. This quantity is used in Ref.~\cite{Szarek2007} to describe the stabilization of molecules through covalent bonds and van der Waals bonds. (It is also used in Ref.~\cite{Szarek2009} recently.)  It is defined in the framework of the Regional DFT and the Rigged QED\cite{Tachibana1999,Tachibana2001,Tachibana2002a,Tachibana2002b,Tachibana2003,Tachibana2004,Tachibana2005,Szarek2008} and can be calculated from the electronic stress tensor density $\overleftrightarrow{\tau}^{S}(\vec{r})$ whose components are given by
\begin{eqnarray} 
\tau^{Skl}(\vec{r})=\frac{\hbar^{2}}{4m}\sum_{i}\nu_{i}
\Bigg[\psi^{*}_{i}(\vec{r})\frac{\partial^{2}\psi_{i}(\vec{r})}{
\partial{x^{k}}\partial{x^{l}}}-\frac{\partial{\psi^{*}_{i}}
(\vec{r})}{\partial{x^{k}}}\frac{\partial{\psi_{i}}
(\vec{r})}{\partial{x^{l}}} 
+\frac{\partial^{2}{\psi^{*}_{i}(\vec{r})}}{\partial{x^
{k}}\partial{x^{l}}}\psi_{i}(\vec{r})-\frac{\partial{\psi
^{*}_{i}}(\vec{r})}{\partial{x^{l}}}\frac{\partial{\psi_
{i}}(\vec{r})}{\partial{x^{k}}}\Bigg],
\end{eqnarray}
where $\{k, l\} = \{1, 2, 3\}$, $m$ is the electron mass,
and $\nu_i$ and $\psi_i(\vec{r})$ is
the occupation number and natural orbital of the $i$th state, respectively. Taking a trace of $\overleftrightarrow{\tau}^{S}(\vec{r})$ gives energy density of the quantum system at each point in space. The energy density $\varepsilon_\tau^S(\vec{r})$ is given by 
\begin{eqnarray}
\varepsilon_\tau^S(\vec{r}) = \frac{1}{2} \sum_{k=1}^3 \tau^{Skk}(\vec{r}).
\end{eqnarray}

Now, we can define the interaction energy density. Suppose that the system under consideration formally consists of two parts A and B and that the whole system has the energy distribution as $\varepsilon_{\tau, {\rm AB}}^S(\vec{r})$. When the parts A and B are considered separately, they have the energy distribution $\varepsilon_{\tau, {\rm A}}^S(\vec{r})$ and $\varepsilon_{\tau, {\rm B}}^S(\vec{r})$ respectively and $\varepsilon_{\tau, {\rm A}}^S(\vec{r}) + \varepsilon_{\tau, {\rm B}}^S(\vec{r}) \neq \varepsilon_{\tau, {\rm AB}}^S(\vec{r})$. The difference stems from stabilization or destabilization due to the reaction between A and B at each point in space and we call it the interaction energy density $\Delta \varepsilon_\tau^S(\vec{r})$. Namely, 
\begin{eqnarray}
\Delta \varepsilon_\tau^S(\vec{r}) = \varepsilon_{\tau, {\rm AB}}^S(\vec{r}) - \left\{ \varepsilon_{\tau, {\rm A}}^S(\vec{r}) + \varepsilon_{\tau, {\rm B}}^S(\vec{r})  \right\}.
\label{eq:ene_int}
\end{eqnarray}
The region with negative $\Delta \varepsilon_\tau^S(\vec{r})$ corresponds to the stabilized region and the positive region denotes the destabilized region. 

We also use conventional electron density difference $\Delta n(\vec{r})$ for the later discussion as defined below.
\begin{eqnarray}
\Delta n_{\rm AB}(\vec{r}) = n_{{\rm AB}}(\vec{r}) - \left\{ n_{\rm A}(\vec{r}) + n_{\rm B}(\vec{r})  \right\},
\label{eq:num_diff}
\end{eqnarray}
where $n(\vec{r})$ is the ordinary electron density at $\vec{r}$.

As shown in Ref.~\cite{Szarek2007}, in covalent bonding, the larger stabilization accompanies the larger atomic population (the region with negative $\Delta \varepsilon_\tau^S(\vec{r})$ has positive $\Delta n(\vec{r})$). However, note that this is not the case for van der Waals bonds, in which the stabilized region corresponds to the $decrease$ in the electron density \cite{Szarek2007} (the region with negative $\Delta \varepsilon_\tau^S(\vec{r})$ has negative $\Delta n(\vec{r})$).

We use Molecular Regional DFT (MRDFT) package \cite{MRDFTv3} to compute these quantities. 

\section{Results and discussion} \label{sec:results}

\subsection{Reaction with water} \label{sec:water}
In this section, we study the reaction of BCl$_3$ with water. Although the hydrolysis of BCl$_3$, ${\rm BCl_3(g)+ 3\,H_2O(l) \rightarrow B(OH)_3(aq) + 3\, HCl(aq)}$, is a well-known textbook-level reaction, the situation we consider here is slightly different. In our case, since H$_2$O is much less (ppm level) than BCl$_3$, a relevant reaction would be ${\rm BCl_3 + H_2O \rightarrow BCl_2(OH) + HCl}$. 

Then, we start by examining how H$_2$O approach BCl$_3$ and form a complex Cl$_3$B---OH$_2$. We take the distance between B and O as a parameter, $D$, and for several values of  $D$, we calculate the optimized configuration of the other atoms. 

The obtained structures are shown in Fig.~\ref{fig:BCl3_H2O_dist}. We can regard them to be the snapshots of the continuous reaction process. While $D$ is between {4.8\,\AA} and {4.0\,\AA}, we find that H$_2$O approach with a small angle from the BCl$_3$ plane. In Fig.~\ref{fig:BCl3_H2O_ene}, we show relative energy and charge transfer as functions of $D$. The charge transfer is calculated from the Mulliken charge. We obtain the structure of the stable reactant complex Cl$_3$B---OH$_2$ as shown in Fig.~\ref{fig:BCl3_H2O_RC} at $D=1.693$\,\AA. Below, we refer to this structure as ``RC". This geometry is in good agreement with the one in the literature \cite{Frenking1997,Rowsell1999}. We find that when H$_2$O approaches BCl$_3$ from an infinite distance, there is no energy barrier and total energy is stabilized by 0.2607\,eV. Also note that the reaction proceeds as charge is transferred from BCl$_3$ to water (electrons from water to BCl$_3$).

Next, we search a reaction path from RC until the detachment of HCl. The energy along a certain intrinsic reaction coordinate (IRC) is plotted in Fig.~\ref{fig:BCl3_H2O_IRC}. Some intermediate structures are also shown. In particular, the one labeled 1 is RC and 3 is the transition state (TS) respectively. This is reorganized in Fig.~\ref{fig:BCl3_H2O_path} and Table \ref{tbl:BCl3_H2O_ene} which show the reaction pathway and relative energy. The activation energy is found to be 0.7300\,eV. By detaching HCl, the system stabilizes by 0.6799\,eV from RC. In the final step, there is energy increment of 0.0990\,eV
which corresponds to the strength of the hydrogen bond between H in HCl and O in BCl$_2$(OH) (the distance between H and O is 2.054\,\AA). 
However, since it is rather small, BCl$_2$(OH) and HCl are expected to be separated away in the gas phase.

Now, we analyze this reaction by using interaction energy density $\Delta \varepsilon_\tau^S(\vec{r})$ and electron density difference $\Delta n(\vec{r})$ as introduced in Sec.~\ref{sec:calc}. We first examine the process from ${\rm BCl_3 + H_2O}$ to RC, namely the reactant complex formation. This is shown in Fig.~\ref{fig:BCl3_H2O_int1}. From $\Delta n(\vec{r})$ of panels (a) and (b), we see that BCl$_3$ and H$_2$O are polarized by the existence of the other molecule when they are separated. Then they attract each other by  electrostatic interaction. After they approach closely as panel (c), electrons move from H$_2$O to BCl$_3$ rapidly as shown by the expansion of pink regions around BCl$_3$ in panels (d), (e) and (f). This is consistent with the charge transfer plot in Fig.~\ref{fig:BCl3_H2O_ene}, which shows steepening trend for $D \lesssim 3.5$\,\AA. 
As for $\Delta \varepsilon_\tau^S(\vec{r})$, general feature we notice is that positive $\Delta n(\vec{r})$ region (colored in pink) roughly corresponds to negative $\Delta \varepsilon_\tau^S(\vec{r})$ region (colored in blue) and vice versa. This indicates that each part in the system is stabilized by the increase in the electron density just as in the usual covalent bonding. 

The process from RC to ${\rm BCl_2(OH) \cdot HCl}$ via TS is next analyzed and results are shown in Fig.~\ref{fig:BCl3_H2O_int2}. We again see the correspondence between positive $\Delta n(\vec{r})$ region and negative $\Delta \varepsilon_\tau^S(\vec{r})$ region. Since we partition the system into BCl$_2$(OH) and HCl for calculating $\Delta n(\vec{r})$ and $\Delta \varepsilon_\tau^S(\vec{r})$, it is easier to see the process backward from the panel (g). We see that red destabilized region expands from panel (g) to (a).
This destabilized region is especially large in TS (panel (c)) around H$_2$O. Although this partitioning is not so well defined in (a) and (b), we see destabilized region around H$_2$O too. This is consistent with the energy level relation that RC has higher energy than ${\rm BCl_2(OH) \cdot HCl}$.

\subsection{Reaction with iron hydroxide}

In this section, we study the reaction of BCl$_3$ with Fe(OH)$_3$. Fe(OH)$_3$ is our model of iron impurity and we investigate how this can react with BCl$_3$ molecules to produce HCl. It should be mentioned that Fe(OH)$_3$ has lower energy than FeO(OH)$\cdot$H$_2$O. As shown in Table \ref{tbl:rel_ene}, since sextet is the most stable, we adopt this state in the following calculation.

The entire reaction path we have searched is shown in Fig.~\ref{fig:Fe_BCl3_pathall}. Table \ref{tbl:Fe_BCl3_pathall} shows relative energy for each step. For convenience, we split this path into four as in Fig.~\ref{fig:Fe_BCl3_path} (a)-(d) showing structures for each step. We will give detailed description for each of them below.

The earliest stages in the reaction path are shown in Fig.~\ref{fig:Fe_BCl3_path} (a). First, it is shown that BCl$_3$ approaches Fe(OH)$_3$ with some energy stabilization forming a reaction complex S$_2$. A possible simple path to detach HCl from S$_2$ is shown as ${\rm S_2 \rightarrow TS_1 \rightarrow S_3}$. This is just like the one we have seen in the reaction between BCl$_3$ and H$_2$O described in Sec.~\ref{sec:water}. We need the activation energy of 1.1653\,eV. In the case of BCl$_3$ and Fe(OH)$_3$, however, the reaction complex S$_2$ can turn into more stable structure by cutting a bond B---Cl in the ring consists of B, Cl, O, and Fe atoms and leaving a Fe---Cl bond (${\rm S_2 \rightarrow S_4}$). Since the activation energy is rather high, it is likely that the latter path is taken in the BCl$_3$ gas. Also, as we will see below, the presence of many BCl$_3$ molecules (relative to Fe(OH)$_3$) makes it possible to open more effective paths to detach HCl molecules.

Now, let us consider the case that another BCl$_3$ approaches S$_4$. One more bond between BCl$_3$ and OH is formed to make S$_5$ and energy is stabilized by 1.1178\,eV. To detach HCl from S$_5$, there are two reaction paths as shown in Fig.~\ref{fig:Fe_BCl3_path} (b). One is from S$_5$ to S$_6$ via TS$_{\rm 2a}$ and another is via TS$_{\rm 2b}$. The former is similar to the BCl$_3$+H$_2$O case or the path ${\rm S_2 \rightarrow TS_1 \rightarrow S_3}$ mentioned just above and its activation energy is relatively high, 0.8665\,eV. By contrast, the latter has much lower barrier of 0.15083\,eV. In this path, since HCl is detached from Cl and OH which are connected to different B, the distance between H and Cl is shorter in TS and much less energy is needed to form the bond. 

The path after that is opened in a similar way (Fig.~\ref{fig:Fe_BCl3_path} (c)). Namely, HCl is not likely to be produced from within S$_7$ but detach after a complex with one more BCl$_3$ is formed. This complex is shown as S$_8$. Then HCl is again formed from Cl and OH which are connected to different B with a relatively low activation energy of 0.08919\,eV (${\rm S_8 \rightarrow TS_3 \rightarrow S_9}$). 

So far, we have learned that we can find an energetically more favorable path to detach HCl by attaching BCl$_3$ beforehand. Then, similar path to ${\rm S_2 \rightarrow S_4 \rightarrow S_5}$ may take place for S$_5$ by attaching BCl$_3$. This turns out to be true and we find a path ${\rm S_5 \rightarrow S_{11} \rightarrow S_{12}}$ as shown in Fig.~\ref{fig:Fe_BCl3_path} (d). S$_{11}$ is stabilized by 0.59024\,eV from S$_5$ by forming another Fe---Cl bond. After that, S$_{12}$ is stabilized by 0.8491\,eV from S$_{11}$ by the interaction between BCl$_3$ and the last OH group bonded to Fe.

Finally, how HCl is detached from S$_{12}$ is described. As shown by the path ${\rm S_{12} \rightarrow TS_4 \rightarrow S_{13}}$, Cl and OH which are connected to different B bond to form HCl. This is similar to ${\rm S_{5} \rightarrow TS_{2b} \rightarrow S_{6}}$ and ${\rm S_{8} \rightarrow TS_{3} \rightarrow S_{9}}$. It also has a relatively low activation energy of 0.22614\,eV. We further examine whether one more HCl can be detached from S$_{13}$. However, we cannot find such a path and we find that BCl$_2$OH is detached instead. Thus, the final product is ${\rm Fe(Cl)_2(OBCl_2)(OHBCl_2)}$ shown as S$_{14}$.

To sum up, we have found 
\begin{eqnarray}
\rm 
Fe(OH)_3 + BCl_3 \rightarrow Fe(OH)_2(OBCl_2) + HCl
\end{eqnarray}
has high activation energy and unlikely to occur but there are two reactions with lower energy barriers which produce HCl and more stable Fe-compounds. They can be summarized as 
\begin{eqnarray}
\rm 
Fe(OH)_3 + 3 BCl_3 &\rightarrow& \rm FeCl(OBCl_2)_2(OHBCl_2) + 2HCl 
\end{eqnarray}
and
\begin{eqnarray}
\rm 
Fe(OH)_3 + 3 BCl_3 &\rightarrow& \rm FeCl_2(OBCl_2)(OHBCl_2) + HCl + BCl_2OH 
\end{eqnarray}

In passing, it may be useful to comment on the geometrical structure of each complex in the reaction path. Generally speaking, a four-coordinate complex forms a square-planar or tetrahedral structure. We calculate the skewness of the complex defined as follows \cite{Ito2007}:
\begin{eqnarray}
\sigma = \frac{V - V_{\rm opt}}{V_{\rm opt}}  \label{eq:skew}
\end{eqnarray}
where $V$ is the volume of the tetrahedron defined by the four atoms directly connected to Fe, and $V_{\rm opt}$  is the volume of a regular tetrahedron which has the common circumsphere to that tetrahedron. If $\sigma=0$, the tetrahedron is regular and if $\sigma=1$, it is square planar. The result is shown in Table \ref{tbl:Fe_BCl3_pathall}. This result shows that the complexes we have dealt with are very close to regular tetrahedrons. 

The analyses using $\Delta n(\vec{r})$ and $\Delta \varepsilon_\tau^S(\vec{r})$ for some parts of the reaction path are shown in Figs.~\ref{fig:Fe_BCl3_int2}-\ref{fig:Fe_BCl3_int1}. As is the cases which are examined in Sec.~\ref{sec:water}, positive $\Delta n(\vec{r})$ region corresponds to negative $\Delta \varepsilon_\tau^S(\vec{r})$ region in general. Fig.~\ref{fig:Fe_BCl3_int2} shows the process of BCl$_3$ approaching a complex Fe(Cl)(OH)$_2$(BCl$_2$OH) to form a reaction complex and Fig.~\ref{fig:Fe_BCl3_int3} shows the process of detaching HCl from the reaction complex. They are respectively similar to what we have seen in Figs.~\ref{fig:BCl3_H2O_int1} and \ref{fig:BCl3_H2O_int2}. Namely, we see blue stabilized region grows between BCl$_3$ and Fe(Cl)(OH)$_2$(BCl$_2$OH) as the process proceeds from (a) to (e) in Fig.~\ref{fig:Fe_BCl3_int2} and red destabilized region develops in the direction from (e) to (a) in Fig.~\ref{fig:Fe_BCl3_int3}.
We show in Fig.~\ref{fig:Fe_BCl3_int1} one more example of $\Delta \varepsilon_\tau^S(\vec{r})$ for the process of detaching HCl from the reaction complex. When we look Fig.~\ref{fig:Fe_BCl3_int1} in the direction from panel (f) to (a), we see that red destabilized region grows, as is the case of Fig.~\ref{fig:Fe_BCl3_int3}.

\section{Conclusion} \label{sec:conclusion}
We have investigated a reaction of BCl$_3$ with Fe(OH)$_3$ by ab initio quantum chemical calculation as one of the simplest models for a reaction of iron impurities in BCl$_3$ gas. We have found that compounds such as ${\rm Fe(Cl)(OBCl_2)_2(OHBCl_2)}$ and ${\rm Fe(Cl)_2(OBCl_2)(OHBCl_2)}$ are formed while producing HCl. The reaction paths to them are examined in detail and their activation energy is found to be relatively low due to the formation of a Fe-complex coordinated by several BCl$_3$ before detaching HCl. We have also examined a reaction with a single H$_2$O molecule (remember that H$_2$O is rare in the BCl$_3$ gas) and have found that it has high activation energy. Such difference in energy barriers indicates that it is more likely that the observed HCl originates from the reaction of BCl$_3$ with iron impurities rather than from the reaction with H$_2$O. 

We have also analyzed the stabilization mechanism of these paths using newly-developed interaction
energy density $\Delta \varepsilon_\tau^S(\vec{r})$ in our laboratory derived from electronic stress tensor in the framework of the Regional DFT and Rigged QED. We have compared this with electron density difference $\Delta n(\vec{r})$. We have found correspondence between positive (negative) $\Delta n(\vec{r})$ region and negative (positive) $\Delta \varepsilon_\tau^S(\vec{r})$ region in general. This indicates a covalent bond that a bond is stabilized by the increase in electron density. We believe this interaction energy density is very useful to analyze and visualize how and in which part of molecules are (de-)stabilized during the chemical process. Integrating the interaction energy density over some region would give good quantitative measure of stabilization. This will be investigated in our future work.

Although it is too early to conclude that the reaction paths we have shown are realized in the BCl$_3$ gas in the ductwork, it is reasonable to imagine iron impurities play some role in producing HCl. More detailed modeling of iron impurities in future would give us more hints for this issue.


\newpage
\begin{table}
\begin{center}
\caption{Relative energy along the pathway of the reaction of BCl$_3$ with water shown in Fig.~\ref{fig:BCl3_H2O_path}.}
\label{tbl:BCl3_H2O_ene}
\bigskip
\begin{tabular}{|c|c|}
\hline
 & Relative energy (eV)  \\
 \hline 
 \hline
BCl$_3$+H$_2$O & 0 \\
RC & $-$0.26065 \\
TS &  0.36501 \\
${\rm BCl_2(OH)\cdot HCl}$ & $-$0.94056 \\
BCl$_2$(OH) + HCl & $-$0.84152 \\
  \hline
\end{tabular}
\end{center}
\end{table}

\begin{table}
\begin{center}
\caption{Relative energy in units of eV for the optimized structure of several spin states of Fe(OH)$_3$ and FeO(OH)$\cdot$H$_2$O. We take sextet of Fe(OH)$_3$ which has the lowest energy as the reference point.}
\label{tbl:rel_ene}
\bigskip
\begin{tabular}{|c|c|c|c|}
\hline
 & Quartet & Sextet & Octet \\
 \hline
 Fe(OH)$_3$ & 0.01366 & 0 & 5.48862 \\
 FeO(OH)$\cdot$H$_2$O & 0.04452  & 0.04936  & 0.17616 \\
 \hline
\end{tabular}
\end{center}
\end{table}

\begin{table}
\begin{center}
\caption{Relative energy and skewness of tetrahedron (Eq.~\eqref{eq:skew}) along the pathway of the reaction of BCl$_3$ with Fe(OH)$_3$ shown in Fig.~\ref{fig:Fe_BCl3_pathall}.}
\label{tbl:Fe_BCl3_pathall}
\bigskip
\begin{tabular}{|c|c|c|}
\hline
 & Relative energy (eV) &  Skewness of tetrahedron $\sigma$ \\
 \hline 
 \hline
 S$_1$ &  0 & (3-coordinate)  \\
 S$_2$ & $-$1.83438 & 0.0990   \\
 TS$_1$ & $-$0.66899 & 0.0990 \\
 S$_3$ & $-$1.70663& 0.1013   \\  
  S$_4$ & $-$2.36146& 0.0961  \\ 
 S$_5$ & $-$3.47929 & 0.0324  \\ 
 TS$_{\rm 2a}$ & $-$2.61284  & 0.0570 \\
 TS$_{\rm 2b}$ & $-$3.32846 & 0.0355  \\ 
 S$_6$ & $-$3.54107 & 0.0291  \\ 
 S$_7$ & $-$3.41744 & 0.0801  \\
 S$_8$ & $-$4.27892 & 0.1591  \\ 
 TS$_3$ & $-$4.18973 & 0.0723 \\
 S$_9$ & $-$4.57131 & 0.0515 \\
 S$_{10}$ & $-$4.47948 & 0.0549 \\
 S$_{11}$ & $-$3.88924 & (5-coordinate) \\
 S$_{12}$ & $-$4.73834 & (5-coordinate) \\
 TS$_4$ & $-$4.51220 &  (5-coordinate) \\
  S$_{13}$ & $-$4.72870 & 0.0364 \\
  S$_{14}$ & $-$4.49319 & 0.0492\\
  \hline
\end{tabular}
\end{center}
\end{table}

\ifFIG
\newpage
\begin{figure}
\begin{center}
\includegraphics[scale=0.7]{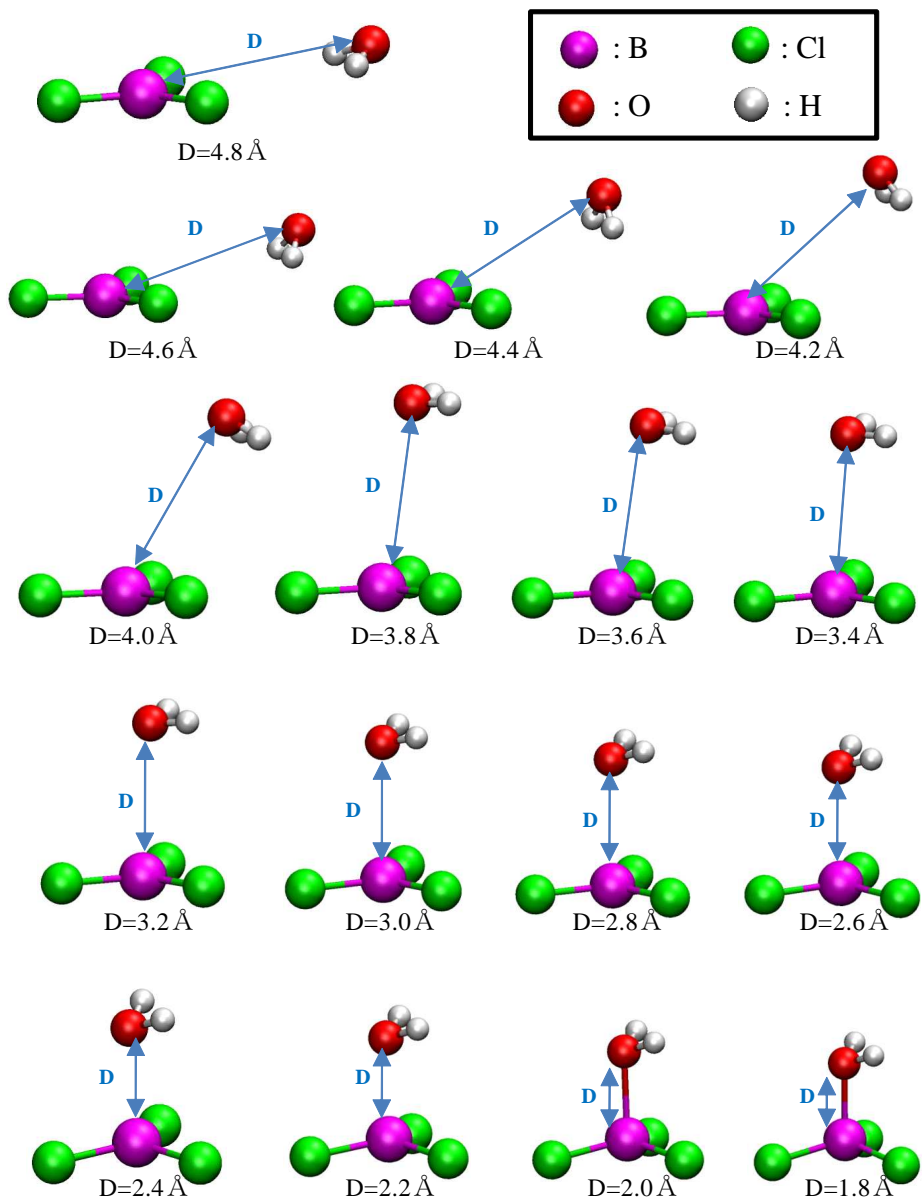}
\caption{Optimized configurations of BCl$_3$ and H$_2$O at the given B---O distance ($D$).}
\label{fig:BCl3_H2O_dist}
\end{center}
\end{figure}

\begin{figure}
\begin{center}
\includegraphics[scale=0.7]{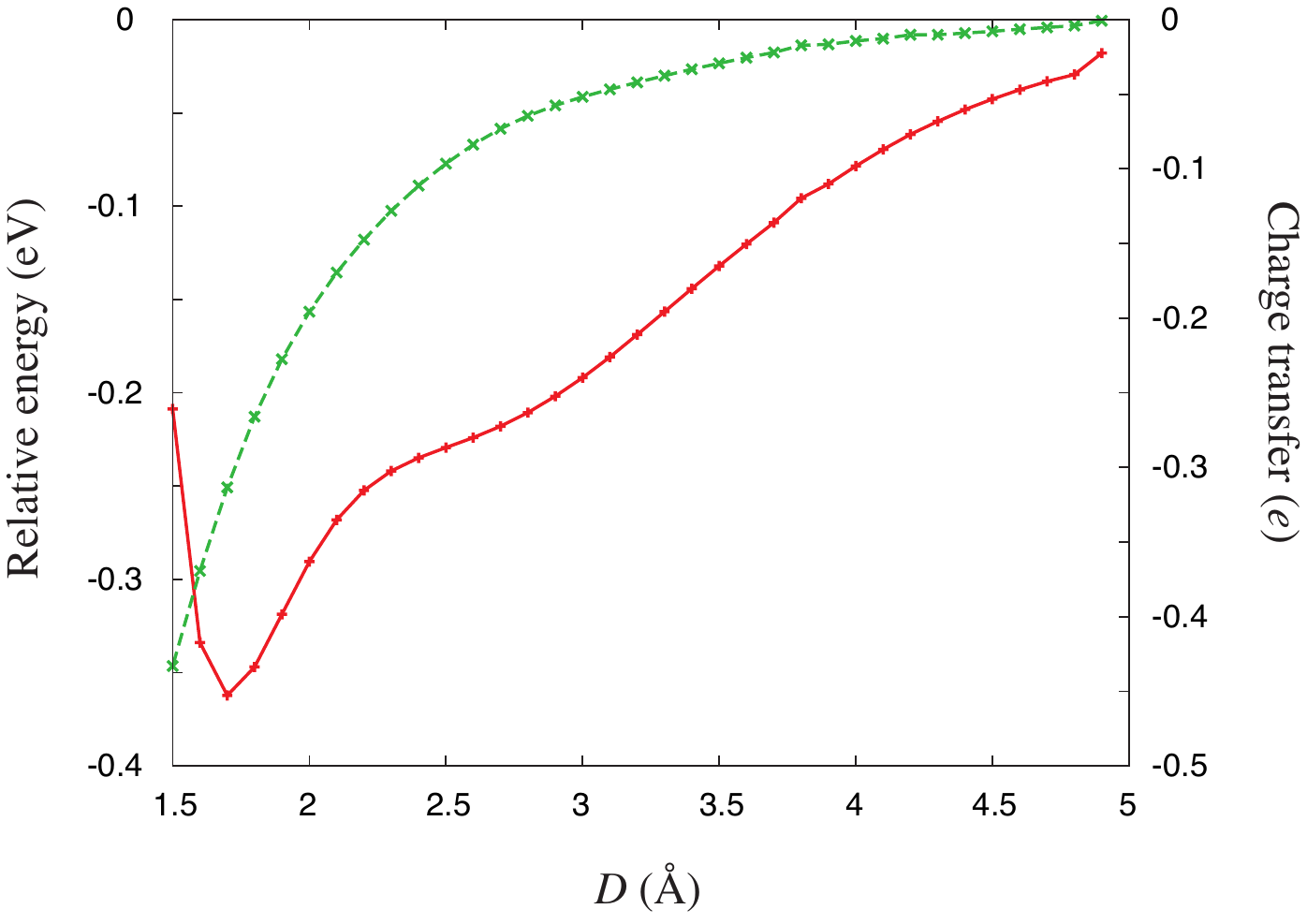}
\caption{The plots of relative energy (solid red line) and charge transfer (dashed green line) vs B---O distance. The reference energy is the energy of structure in which H$_2$O locates at an infinite distance from BCl$_3$. The change in the charge of BCl$_3$ is plotted as the charge transfer on the right axis. Namely, the negative charge transfer indicates that electrons moves in to BCl$_3$ from H$_2$O.} 
\label{fig:BCl3_H2O_ene}
\end{center}
\end{figure}

\begin{figure}
\begin{center}
\includegraphics[scale=0.8]{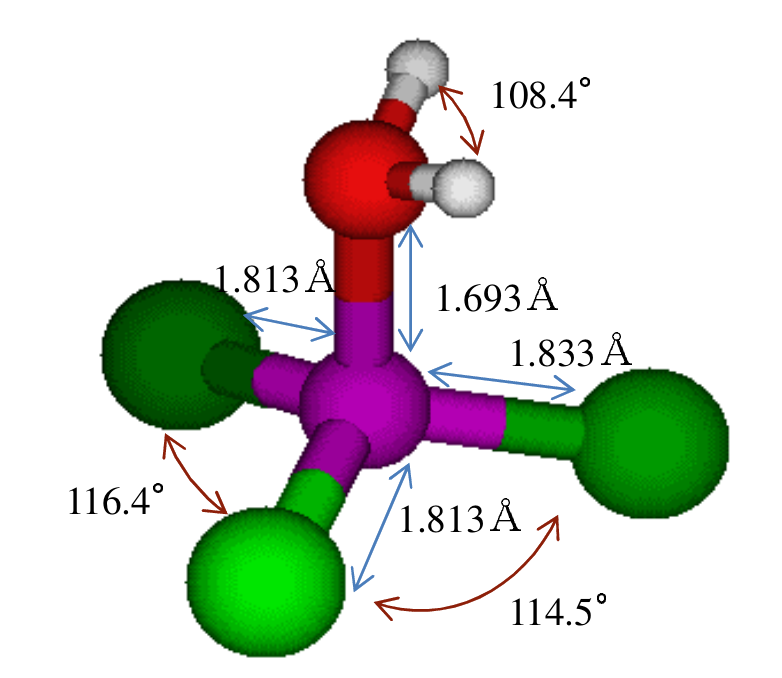}
\caption{The structure of the stable reactant complex (RC), Cl$_3$B---OH$_2$.}
\label{fig:BCl3_H2O_RC}
\end{center}
\end{figure}

\begin{figure}
\begin{center} 
\includegraphics[scale=0.7]{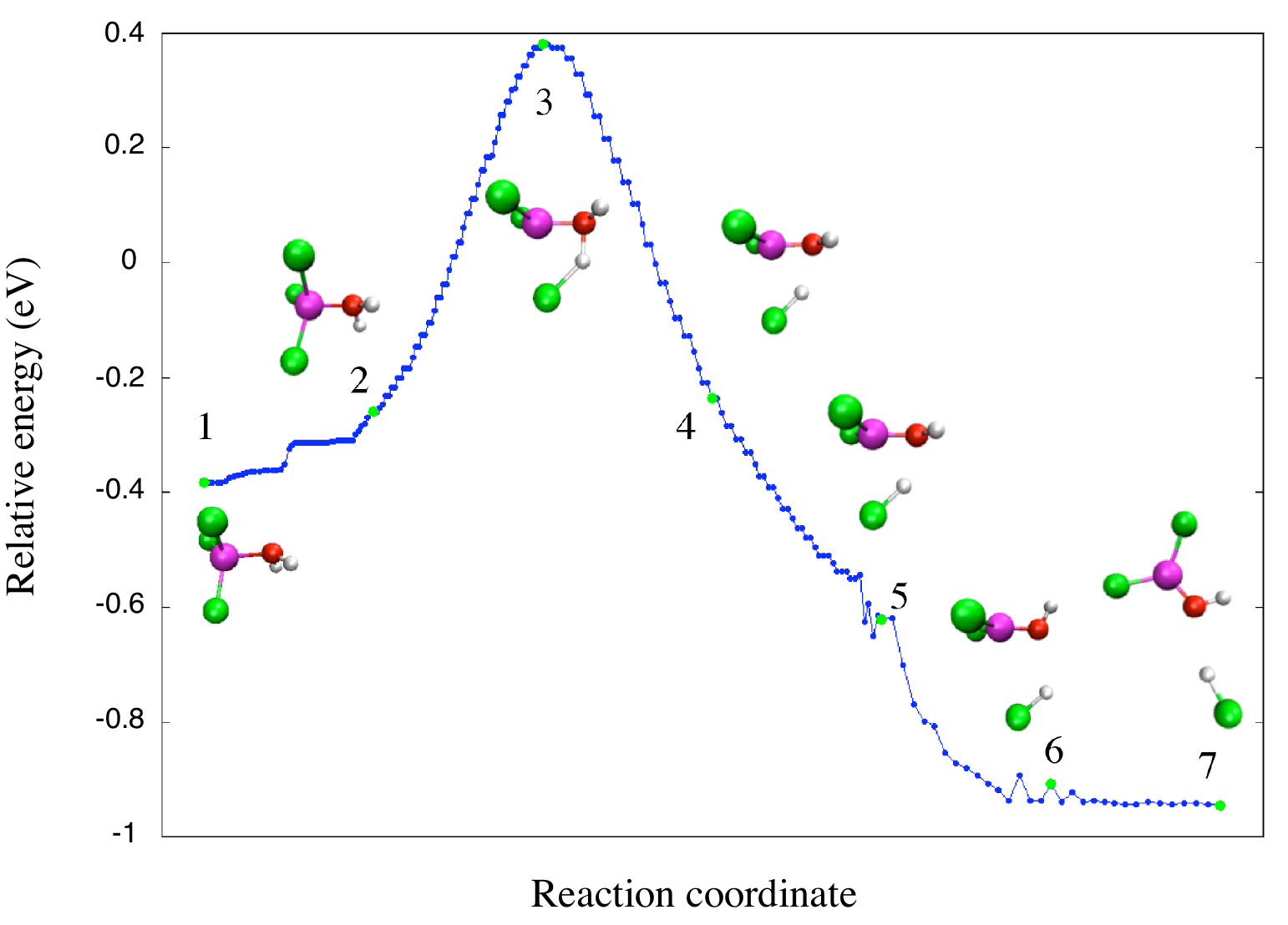}
\caption{Relative energy of the path from Cl$_3$B---OH$_2$ (RC) to ${\rm BCl_2(OH)\cdot HCl}$. The reference energy is the energy of structure in which H$_2$O locates at an infinite distance from BCl$_3$.
}
\label{fig:BCl3_H2O_IRC}
\end{center}
\end{figure}

\begin{figure}
\begin{center}
\includegraphics[scale=0.7]{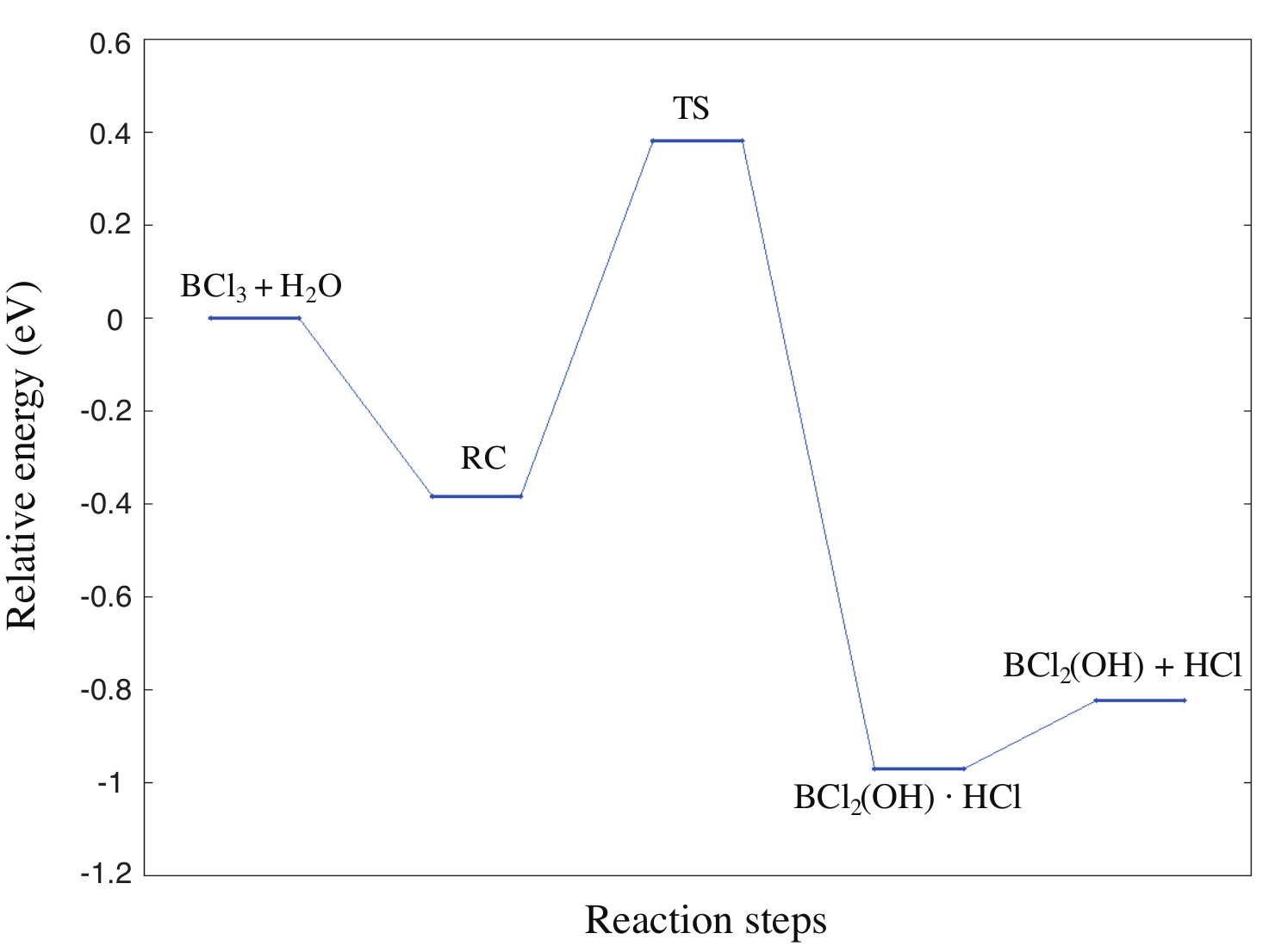}
\caption{Reaction pathway from ${\rm BCl_3 + H_2O}$ to ${\rm BCl_2(OH) + HCl}$.}
\label{fig:BCl3_H2O_path}
\end{center}
\end{figure}

\begin{figure}
\begin{center}
\includegraphics[scale=0.7]{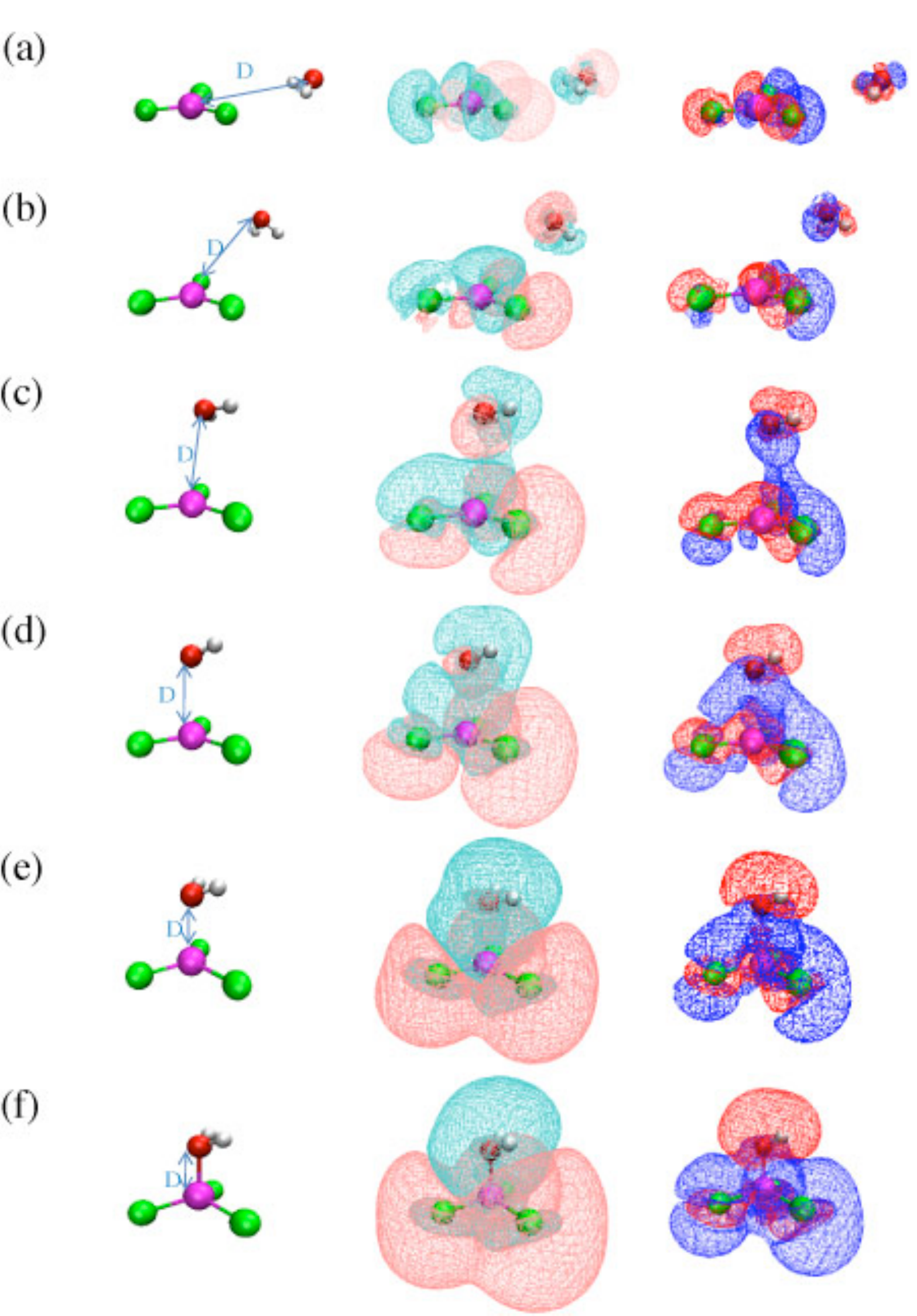}
\caption{Optimized configurations (left column), electron density difference $\Delta n(\vec{r})$ (middle column) and interaction energy density $\Delta \varepsilon_\tau^S(\vec{r})$ (right column) during the process of the Cl$_3$B---OH$_2$ (RC) formation. We partition the system into BCl$_3$ and H$_2$O for calculating $\Delta n(\vec{r})$ and $\Delta \varepsilon_\tau^S(\vec{r})$. For $\Delta n(\vec{r})$, the light blue region has negative value and the pink region has positive value. For $\Delta \varepsilon_\tau^S(\vec{r})$, the blue region has negative value ($i.e.$ stabilized region) and the red region has positive value ($i.e.$ destabilized region).  Each panel shows the B---O distance of (a) 4.8\,\AA, (b) 4.2\,\AA, (c) 3.4\,\AA, (d) 2.7\,\AA, (e) {2.2\,\AA} and (f) {1.693\,\AA} (RC).}
\label{fig:BCl3_H2O_int1}
\end{center}
\end{figure}

\begin{figure}
\begin{center}
\includegraphics[scale=0.55]{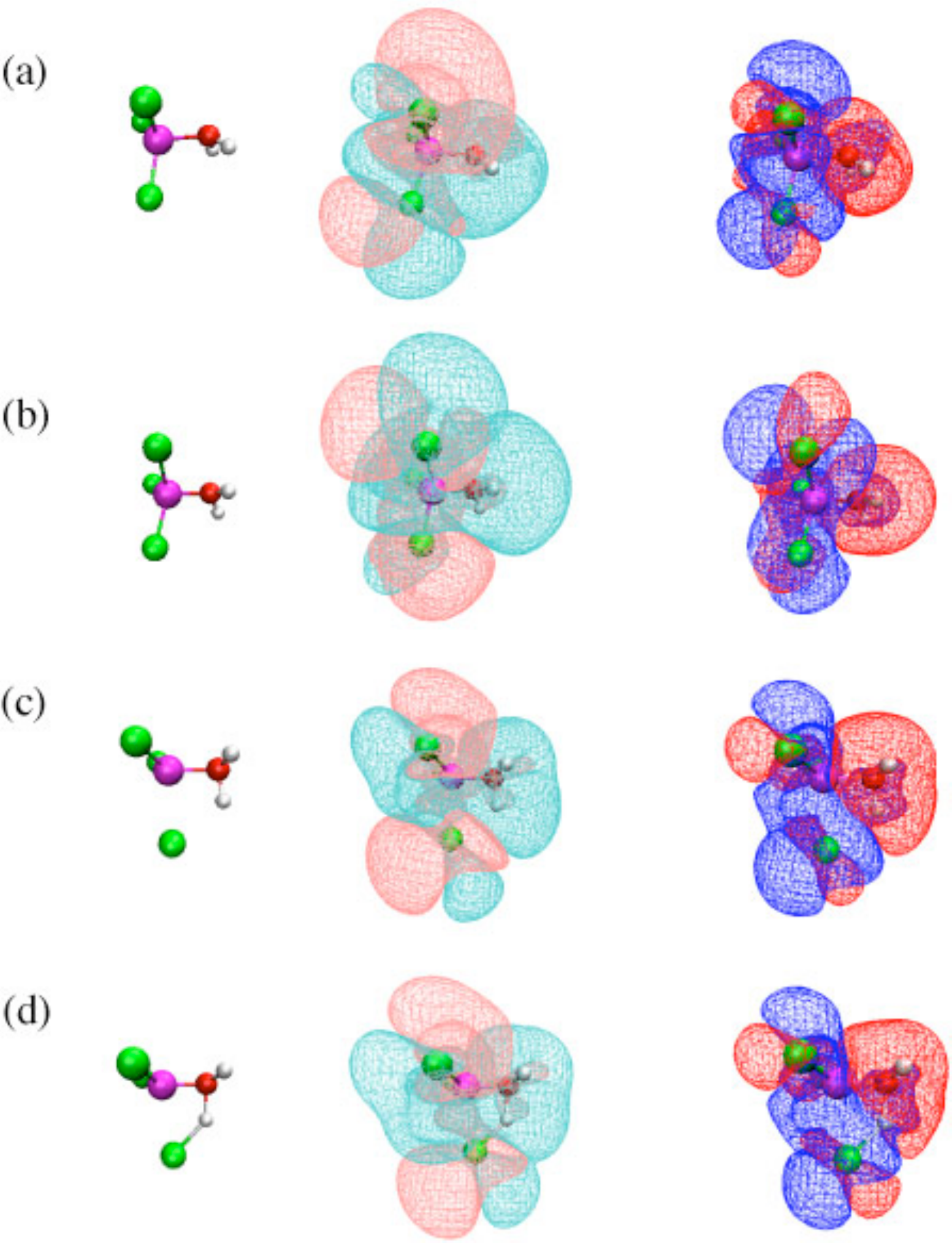}
\includegraphics[scale=0.55]{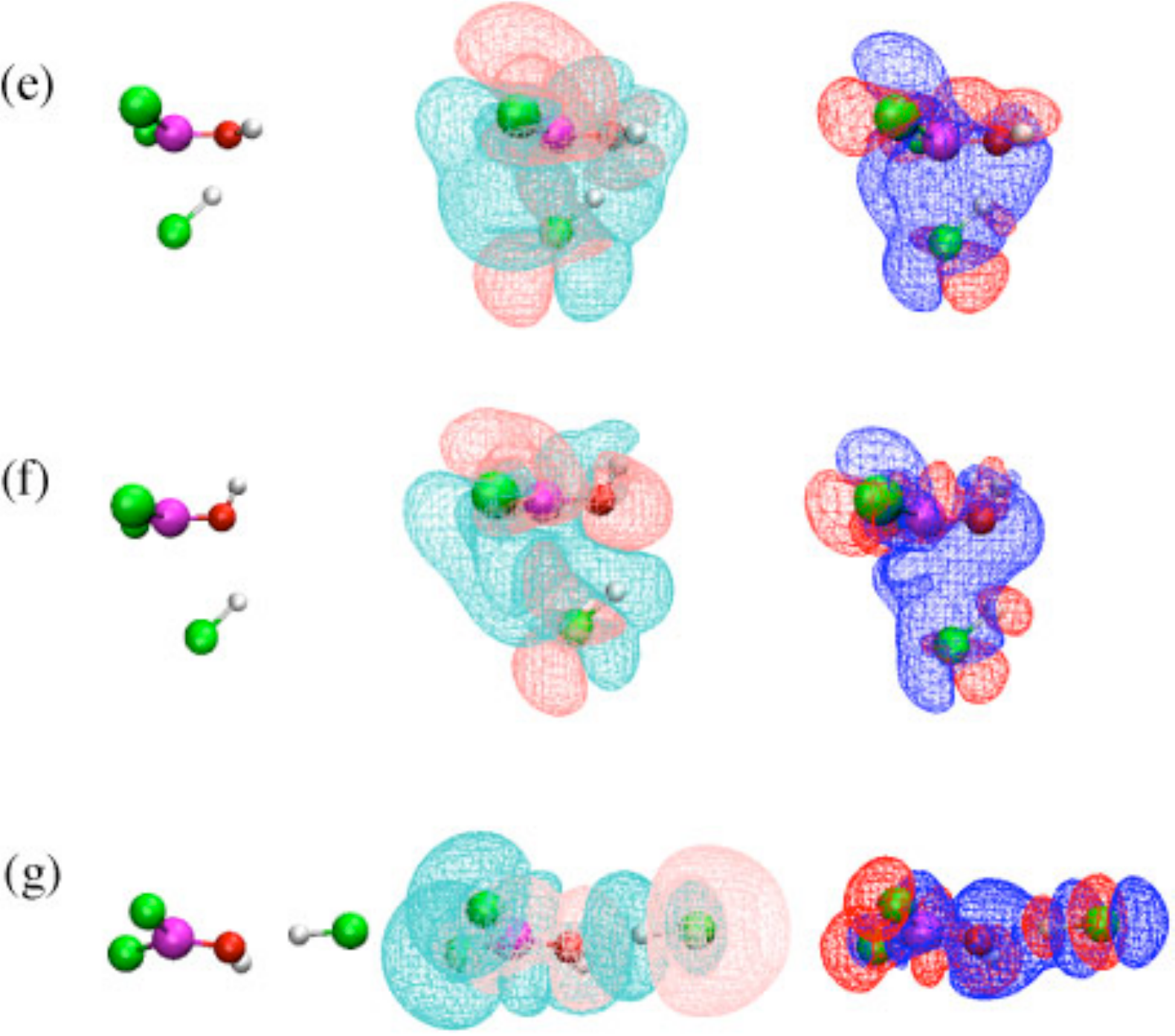}
\caption{Same as Fig.~\ref{fig:BCl3_H2O_int1} for the process from Cl$_3$B---OH$_2$ (RC) to the HCl detachment. We partition the system into BCl$_2$(OH) and HCl for calculating $\Delta n(\vec{r})$ and $\Delta \varepsilon_\tau^S(\vec{r})$. Each panel (a)--(g) corresponds respectively to the step 1--7 indicated in Fig.~\ref{fig:BCl3_H2O_IRC}. The panel (c) describes TS.}
\label{fig:BCl3_H2O_int2}
\end{center}
\end{figure}

\begin{figure}
\begin{center}
\includegraphics[scale=0.7]{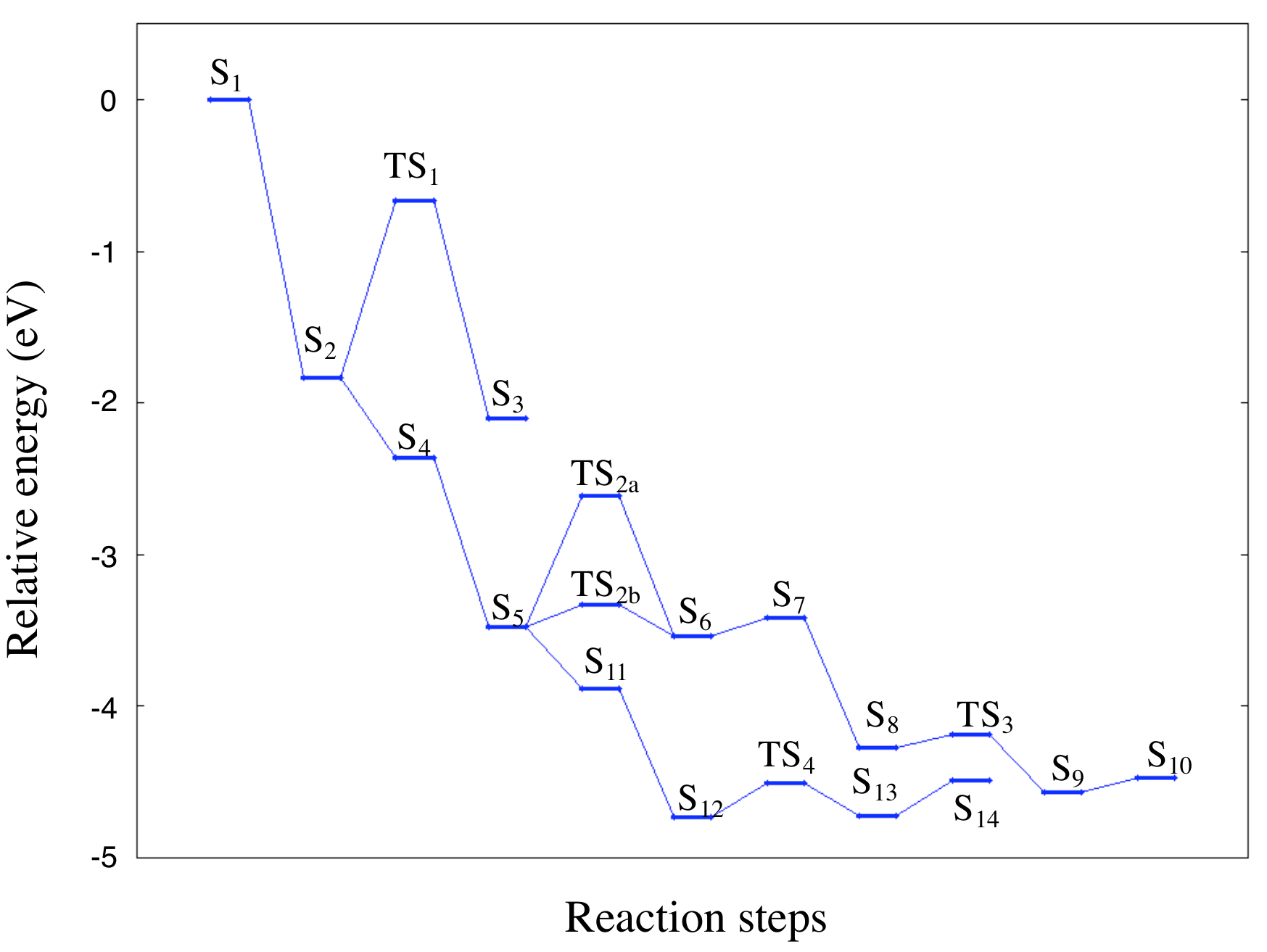}
\caption{The pathway of reaction of Fe(OH)$_3$ with BCl$_3$ toward ${\rm Fe(Cl)(OBCl_2)_2(OHBCl_2)}$ (S$_{10}$) and ${\rm Fe(Cl)_2(OBCl_2)(OHBCl_2)}$ (S$_{14}$).}
\label{fig:Fe_BCl3_pathall}
\end{center}
\end{figure}

\begin{figure*}
\begin{center}
\includegraphics[scale=0.425]{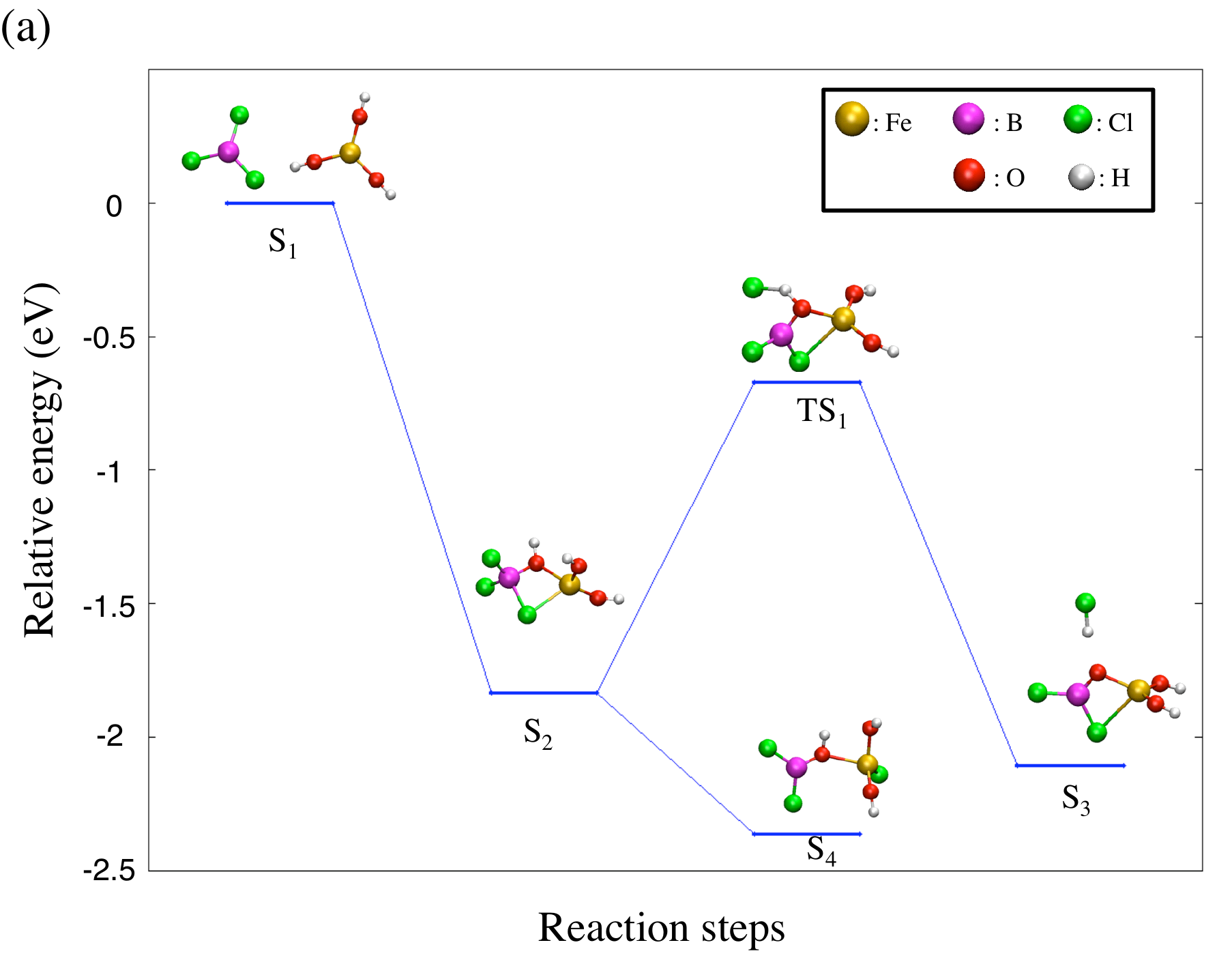}
\hfill
\includegraphics[scale=0.425]{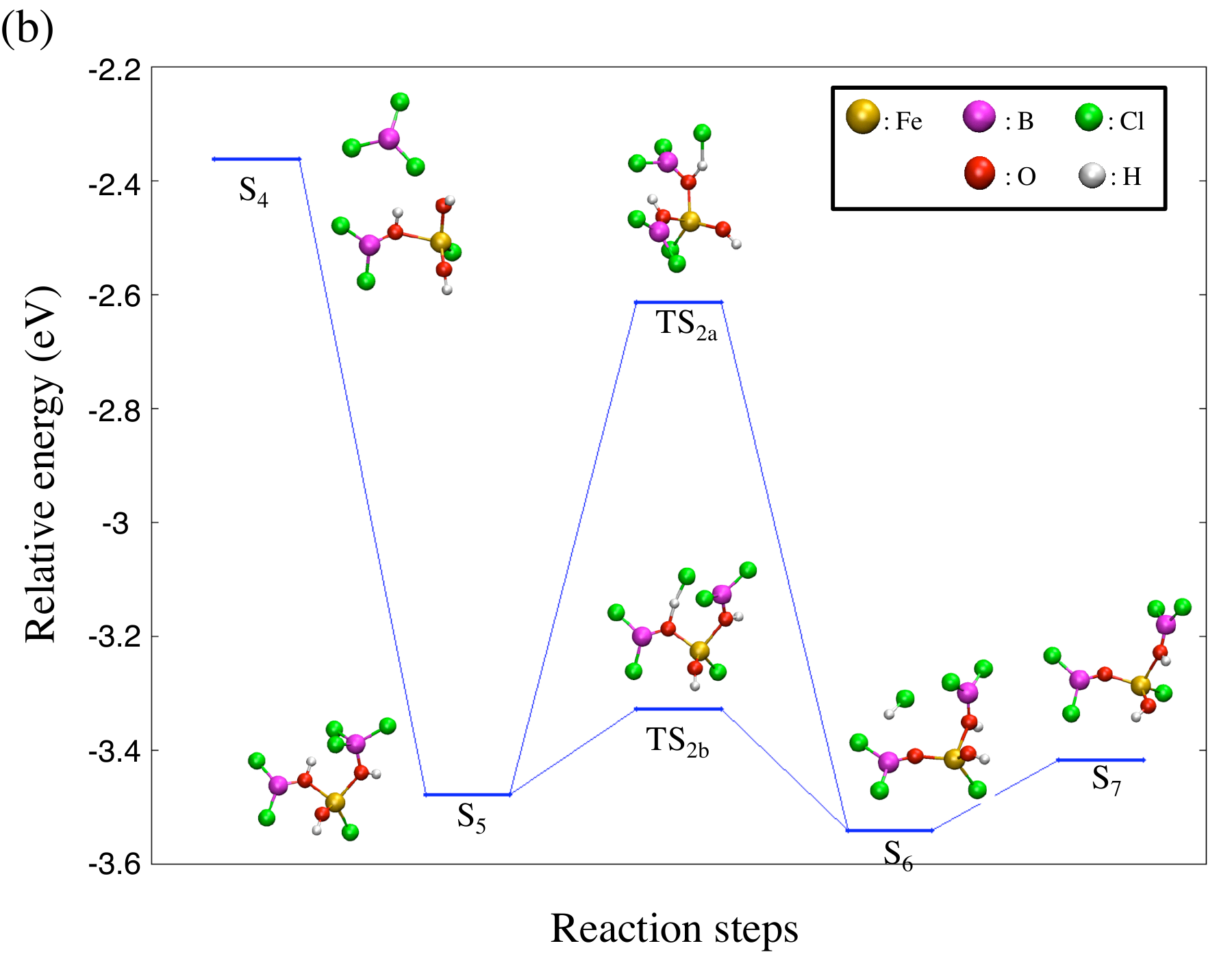} \\
\includegraphics[scale=0.425]{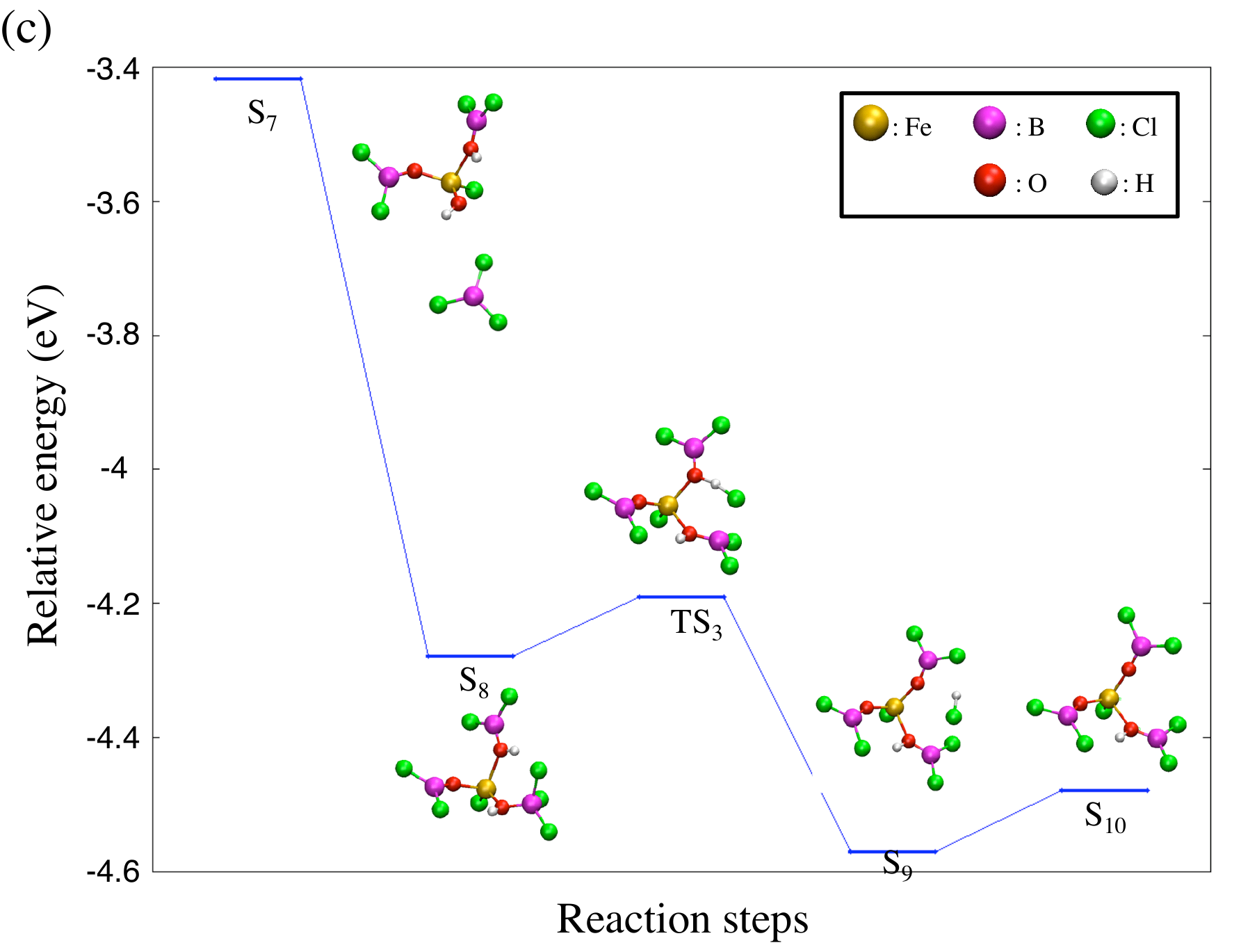}
\hfill
\includegraphics[scale=0.425]{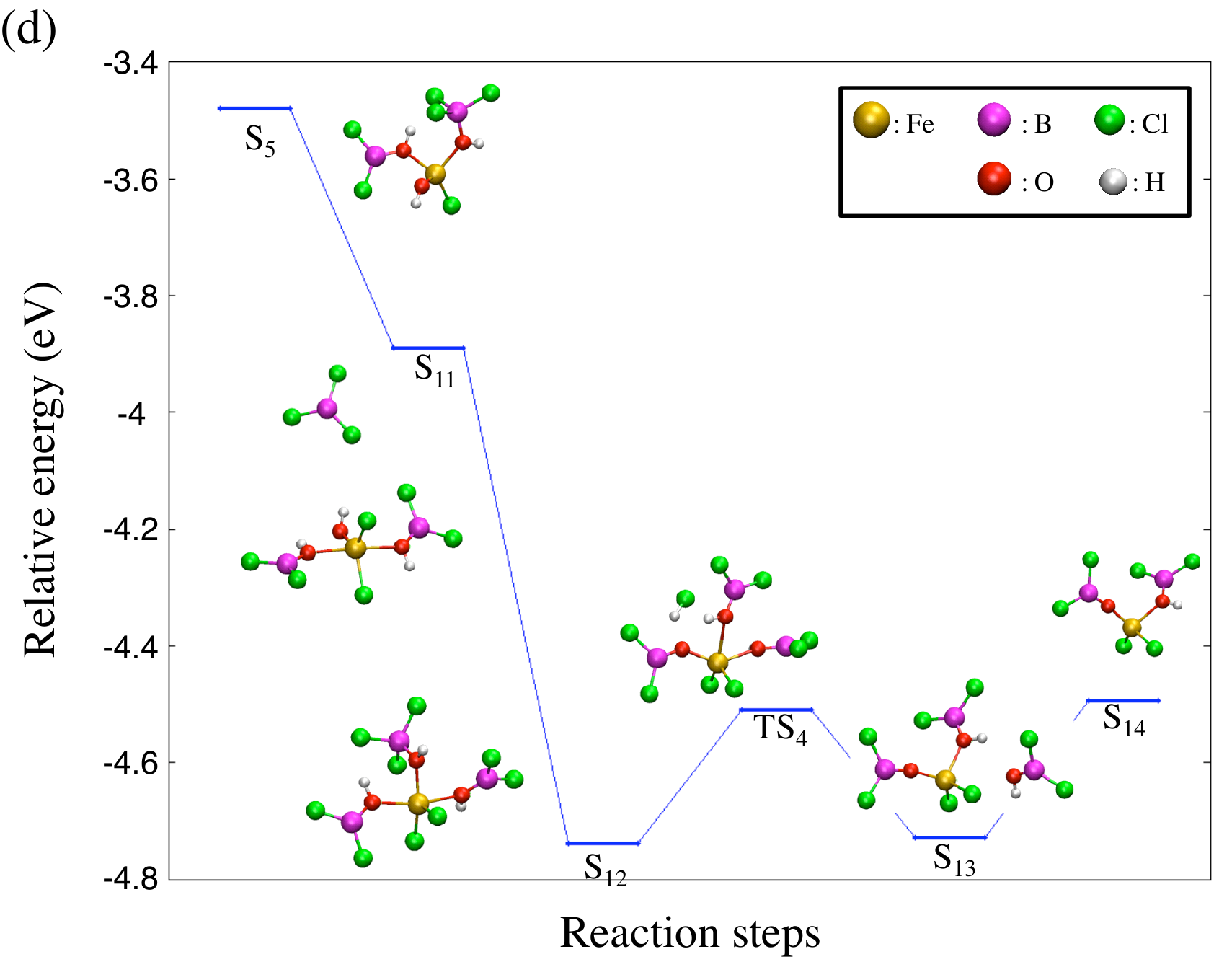}
\caption{The reaction pathways (a) to ${\rm Fe(OH)_2(OBCl_2) \cdot HCl}$ (S$_3$) and to FeCl(OH)$_2$(OHBCl$_2$) (S$_4$), (b) to ${\rm Fe(OH)(Cl)(OBCl_2)(OHBCl_2)}$ (S$_7$), (c) to ${\rm Fe(Cl)(OBCl_2)_2(OHBCl_2)}$ (S$_{10}$), and (d) to ${\rm Fe(Cl)_2(OBCl_2)(OHBCl_2)}$ (S$_{14}$). }
\label{fig:Fe_BCl3_path}
\end{center}
\end{figure*}

\begin{figure}
\begin{center}
\includegraphics[scale=0.7]{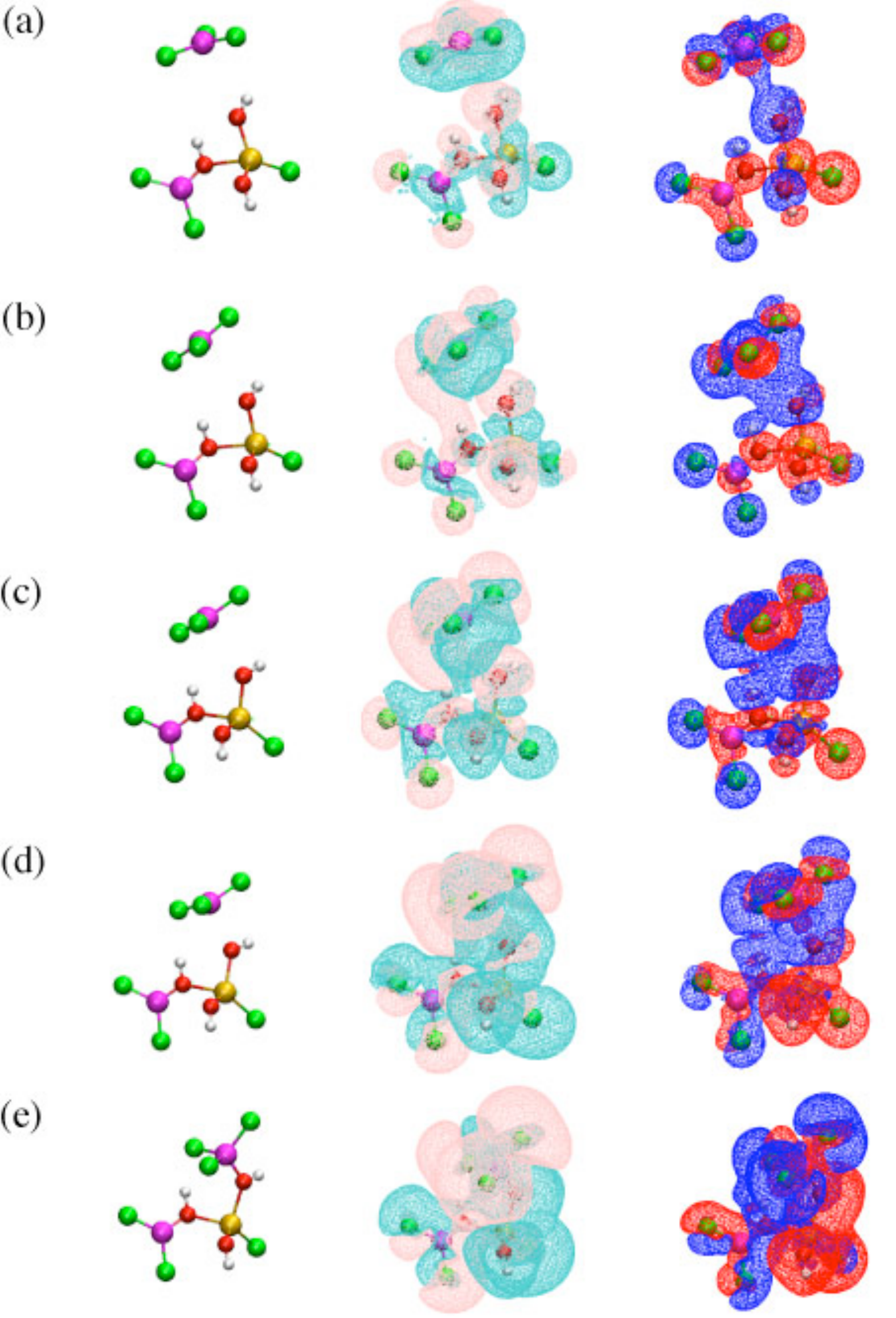}
\caption{Same as Fig.~\ref{fig:BCl3_H2O_int1} for the process ${\rm S_4  \rightarrow S_5}$ (see also Fig.~\ref{fig:Fe_BCl3_path} (b)). We partition the system into Fe(Cl)(OH)$_2$(BCl$_2$OH) and BCl$_3$ for calculating $\Delta n(\vec{r})$ and $\Delta \varepsilon_\tau^S(\vec{r})$. }
\label{fig:Fe_BCl3_int2}
\end{center}
\end{figure}

\begin{figure}
\begin{center}
\includegraphics[scale=0.7]{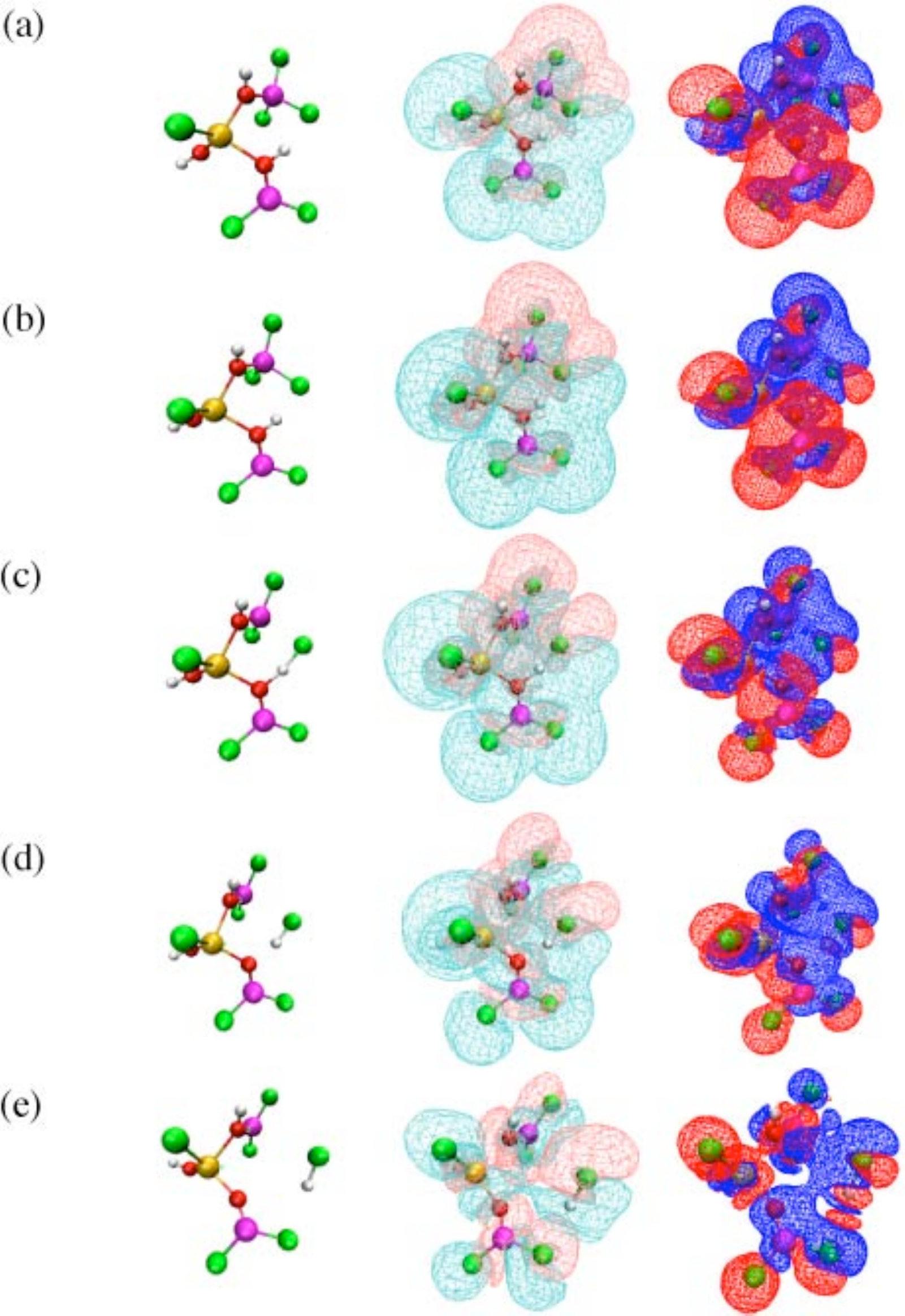}
\caption{Same as Fig.~\ref{fig:BCl3_H2O_int1} for the process ${\rm S_5  \rightarrow TS_{2b} \rightarrow S_6}$ (see also Fig.~\ref{fig:Fe_BCl3_path} (b)). We partition the system into Fe(Cl)(OH)(BCl$_2$O)(BCl$_2$OH) and HCl for calculating $\Delta n(\vec{r})$ and $\Delta \varepsilon_\tau^S(\vec{r})$. TS$_{\rm 2b}$ is denoted by the panel (c).}
\label{fig:Fe_BCl3_int3}
\end{center}
\end{figure}

\begin{figure}
\begin{center}
\includegraphics[scale=0.7]{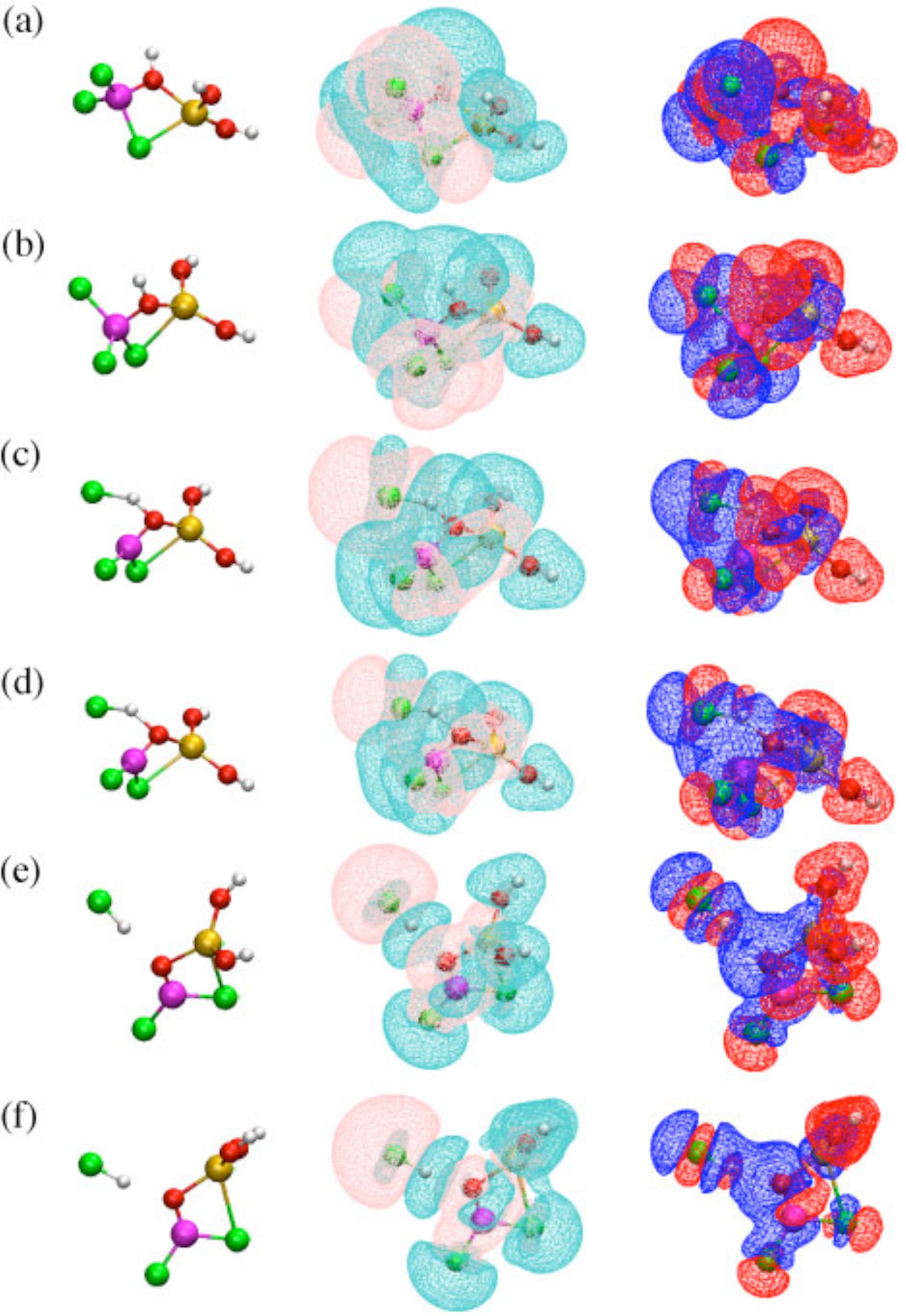}
\caption{Same as Fig.~\ref{fig:BCl3_H2O_int1} for the process ${\rm S_2  \rightarrow TS_1 \rightarrow S_3}$     (see also Fig.~\ref{fig:Fe_BCl3_path} (a)). We partition the system into Fe(OH)$_2$(BCl$_2$O) and HCl for calculating $\Delta n(\vec{r})$ and $\Delta \varepsilon_\tau^S(\vec{r})$. TS$_1$ is denoted by the panel (c).}
\label{fig:Fe_BCl3_int1}
\end{center}
\end{figure}

\fi
\end{document}